\documentclass[10pt,journal,cspaper,compsoc]{IEEEtran}
\usepackage{amsmath,amssymb,amsthm}
\usepackage{graphicx}
\usepackage{verbatim}
\usepackage{multirow,cite}
\usepackage{epsfig}
\usepackage{algorithm,algorithmic}
\usepackage{balance}
\newcommand{\mb}[1]{\ensuremath{\mathbf{#1}}}

\newtheorem{thm}{Theorem}[section]
\newtheorem{lemma}{Lemma}[section]

\usepackage{colortbl}

\def\mb{\mathbf}

\begin{document}
\title{An Enhanced Multiway Sorting Network\\
Based on $n$-Sorters}

\author{Feng Shi, Zhiyuan Yan, and Meghanad Wagh

  \IEEEcompsocitemizethanks{\IEEEcompsocthanksitem Feng Shi, Zhiyuan Yan, and Meghanad Wagh
    are with the Department of ECE,
    Lehigh University, PA 18015, USA. E-mails:\{fes209, yan, mdw0\}@lehigh.edu.}}


%
\IEEEcompsoctitleabstractindextext{
\begin{abstract}
Merging-based sorting networks are an important family of sorting networks. Most merge sorting networks are based on 2-way or multi-way merging algorithms using 2-sorters as basic building blocks.
An alternative is to use $n$-sorters, instead of 2-sorters, as the basic building blocks so as to greatly reduce the number of sorters as well as the latency. Based on a modified Leighton's columnsort algorithm, an $n$-way merging algorithm, referred to as SS-Mk, that uses $n$-sorters as basic building blocks was proposed.
In this work, we first propose a new multiway merging algorithm with $n$-sorters as basic building blocks that merges $n$ sorted lists of $m$ values each in $1+\lceil m/2 \rceil$ stages ($n\le m$). Based on our merging algorithm, we also propose a sorting algorithm, which requires $O(N \log^2 N)$ basic sorters to sort $N$ inputs. While the asymptotic complexity (in terms of the required number of sorters) of our sorting algorithm is the same as the SS-Mk, for wide ranges of $N$, our algorithm requires fewer sorters than the SS-Mk.
Finally, we consider a binary sorting network, where the basic sorter is implemented in threshold logic and scales linearly with the number of inputs, and compare the complexity in terms of the required number of gates. For wide ranges of $N$, our algorithm requires fewer gates than the SS-Mk.
\end{abstract}
}

\maketitle

\section{Introduction}
\IEEEPARstart{S}orting is one important operation in data processing, and hence its efficiency greatly affects the overall performance of a wide variety of applications \cite{knuth1973art,Bat68sorting}.
Sorting networks can achieve high throughput rates by performing operations simultaneously.
These parallel sorting networks have attracted attention of researchers due to increasing hardware speed and decreasing hardware cost. One of the most popular sorting algorithm is called merge-sort algorithm, which performs the sorting in two steps~\cite{Bat68sorting}. First, it divides the input \textbf{list} (a sequence of values) into multiple \textbf{sublists} (a smaller sequence of values) and sorts each sublist simultaneously. Then, the sorted sublists are merged as a single sorted list. The sorting process of sublists can then be decomposed recursively into the sorting and merging of even smaller sublists, which are then merged as a single sorted list. Hence, the merging operation is the key procedure for the decomposition-based sorting approach.
One popular 2-way merging algorithm called odd-even merging~\cite{Bat68sorting} merges two sorted lists (odd and even lists) into one sorted list. In \cite{liszka1992modulo}, a modulo merge sorting was introduced as a generalization of the odd-even merge by dividing the two sorted input lists into multiple sublists with a modulo not limited to 2. Another popular 2-way merging algorithm is bitonic merging algorithm \cite{batcher1990bitonic}. Two sorted lists are first arranged as a bitonic list, which is then converted to obtain a sorted list.
These 2-way merging algorithms employ 2-way merge procedure recursively and have a capability of sorting $N$ values in $O(\log^2 N)$ stages \cite{Bat68sorting}.
In \cite{ajtai19830}, a sorting network, named AKS sorting network, with $O(\log N)$ stages was proposed. However, there is a very large constant in the depth expression, which makes it impractical. Recently, a modular design of high-throughput low-latency sorting units are proposed in \cite{farmahini2013modular}.
However, the basic building block in these 2-way merging algorithm is a 2-sorter, which is simply a $2\times 2$ switching element or comparator as shown in Fig.~\ref{fig:2ksorter}(a).

Instead of using 2-sorters, $n$-sorters can be used as basic building blocks. This was first proposed as a generalization of the Batcher's odd-even merging algorithm~\cite{Lee95multiway}. It was also motivated by the use of $n$-sorters, which sort $n$ ($n\ge 2$) values in unit time \cite{parker1989constructing,beigel1990sorting}. Since large sorters are used as basic building blocks, the number of sorters as well as the latency is expected to be reduced greatly.
An $n$-way merging algorithm was first proposed by Lee and Batcher~\cite{Lee95multiway}, where $n$ is not restricted to 2. A version of the bitonic $n$-way merging algorithm was proposed by Nakatani \textit{et al.} \cite{nakatani1989e1,lee1994sorting}. However, the combining operation in the $n$-way merging algorithms still use 2-sorters as basic building blocks. Leighton proposed an algorithm for sorting $r$ lists of $c$ values each, represented as an $r \times c$ matrix~\cite{leighton1984tight}. This algorithm is a generalization of the odd-even merge-sort and named columnsort, since it merges all sorted columns to obtain a single sorted list in row order. In the original columnsort, no specific operation was provided for sorting columns and no recursive construction of sorting network was provided. In \cite{parker1989constructing}, a modified columnsort algorithm was proposed with sorting networks constructed from $n$-sorters ($n\ge 2$) \cite{liszka1993generalized}. However, a 2-way merge is still used for the merging process.
In~\cite{gao1997sloping}, an $n$-way merging algorithm, named SS-Mk, based on the modified columnsort was proposed with $n$-sorters as basic building blocks, where $n$ is prime. For $n$ sorted lists of $m$ values each, the idea is to sort the $m\times n$ values first in each row and then in slope lines with decreasing slope rates. An improved version of the SS-Mk merge sort, called ISS-Mk, was provided in~\cite{zhao1998efficient}, where $n$ can be any integer. We compare our sorting scheme with the SS-Mk but not the ISS-Mk, because for our interested ranges of $N$, the ISS-Mk requires larger latency due to a large constant.

In this work, we propose an $n$-way merging algorithm, which generalizes the odd-even merge by using $n$-sorters as basic building blocks, where $n$ ($\ge 2$) is prime. Based on this merging algorithm, we also propose a sorting algorithm. For $N = n^{p}$ input values, $p+\lceil n/2 \rceil \times \frac{p(p-1)}{2}$ stages are needed. The complexity of the sorting network is evaluated by the total number of $n$-sorters. The closed-form expression for the number of sorters is also derived.

Instead of 2-sorters, $n$-sorters ($n > 2$) are used as basic blocks in this work. This is because larger sorters have some efficient implementation. For example, for binary sorting in threshold logic, the area of an $n$-sorter scales linearly with the number of inputs $n$, while the latency stays as a constant. Hence, a smaller number of sorters and latency of the whole sorting network can be achieved. However, we cannot use arbitrary large sorters as basic blocks, since larger sorters are more complex and difficult to be implemented. Hence, the benefit of using a larger block diminishes with increasing $n$.
We assume that the size of basic sorter $n \le$ 20 and 10 when evaluating the number of sorters and latency. Our algorithm works for any upper bound on $n$, and one can plug any upper bound on $n$ into our algorithm. Asymptotically, the number of sorters required by our sorting algorithm is on the same order of $O(N\log^2 N)$ as the SS-Mk~\cite{gao1997sloping} for sorting $N$ inputs. Our sorting algorithm requires fewer sorters than the SS-Mk in~\cite{gao1997sloping} in wide ranges of $N$.
For instance, for $n\le 20$, when $N \le 1.46\times 10^4$, our algorithm requires up to 46\% fewer sorters than the SS-Mk. When $1.46 \times 10^4 < N \le 1.3 \times 10^5$, our algorithm has fewer sorters for some segments of $N$'s. When $N > 1.3 \times 10^5$, our algorithm needs more sorters.

The work in this paper is different from previous works~\cite{Lee95multiway,gao1997sloping,zhao1998efficient} in the following aspects:
\begin{itemize}
  \item While the multiway merge \cite{Lee95multiway} uses 2-sorters in the combining network, our proposed $n$-way merging algorithm uses $n$-sorters as basic building blocks. By using larger sorters ($n > 2$), the number of sorters as well as the latency would be reduced greatly.
  \item The merge-based sorting algorithms in \cite{gao1997sloping,zhao1998efficient} are based on the modified columnsort \cite{liszka1993generalized}, which merges sorted columns as a single sorted list in row order. Our $n$-way merge sorting algorithm is a direct generalization of the multiway merge sorting in \cite{Lee95multiway}.
  \item We analyze the performance of our approach by deriving the closed-form expressions of the latency and the number of sorters. We also derive the closed-form expression of the number of sorters for the SS-Mk~\cite{gao1997sloping}, since it was not provided in~\cite{gao1997sloping}. Then we present extensive comparisons between  the latency and the number of sorters required by our approach and the SS-Mk~\cite{gao1997sloping}.
  \item Finally, we show an implementation of a binary sorting network in threshold logic. With an implementation of a large sorter in threshold logic, we compare the performance of sorting networks in terms of the number of gates.
\end{itemize}

The rest of the paper is organized as following. In Sec.~\ref{sec:bg}, we briefly review the background of sorting networks. In Sec.~\ref{sec:kmerge}, we propose a multiway merging algorithm with $n$-sorters as basic blocks. In Sec.~\ref{sec:ksort}, we introduce a multiway sorting algorithm based on the proposed merging algorithm, and show extensive results for the comparison of our sorting algorithm and previous works. In Sec.~\ref{sec:appl}, we focus on a binary sorting network, where basic sorters are implemented by threshold logic and have complexity linear with the input size, and measure the complexity in terms of number of gates. Finally Sec.~\ref{sec:conclusion} presents the conclusion of this work.

\section{Background}
\label{sec:bg}
\begin{figure}[!t]
\centering
\includegraphics[width=6.5cm]{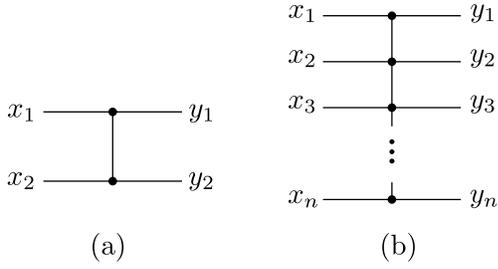}
\caption{(a) 2-sorter ($y_1 \le y_2$); (b) $n$-sorter ($y_1 \le y_2 \le \cdots \le y_n$).}
\label{fig:2ksorter}
\end{figure}

A sorting network is a feedforward network, which gives a sorted list for unsorted inputs. It is composed of two items: \textbf{switching elements (or comparators)} and \textbf{wires}. The depth of a comparator is defined to be the longest length from the inputs of the sorting network to that comparator's outputs. The \textbf{latency} of the sorting network is the maximum depth of all comparators.
The network is \textbf{oblivious} in the sense that the time and location of input and output are fixed ahead of time and not dependent on the values~\cite{Bat68sorting}.
We use the Knuth diagram in \cite{knuth1973art} for easy representation of the sorting networks, where switching elements are denoted by connections on a a set of wires. The inputs enter at one side and sorted values are output at the other side, and what remains is how to arrange the switching elements. The sorting network is measured in two aspects, latency (number of stages) and complexity (number of sorters).
The basic building block used by the odd-even merge~\cite{Bat68sorting} is a 2-by-2 comparator (compare-exchange element). It receives two inputs and outputs the minimum and maximum in an ordered way.
The symbol for a 2-sorter is shown in Fig.~\ref{fig:2ksorter}(a), where $x_i$ and $y_i$ for $i=1,2$ are input and output, respectively.
Similarly, an $n$-sorter is a device sorting $n$ values in unit time. The symbol for an $n$-sorter is shown in Fig.~\ref{fig:2ksorter}(b), where $x_i$ and $y_i$ for $i=1,2,\cdots,n$ are input and output, respectively, and the output satisfies $y_1\le y_2 \le \cdots \le y_n$.
In this work, we denote the sorted values $y_1 \le y_2 \le \cdots \le y_n$ by $\langle y_1, y_2,\cdots, y_n \rangle$ and use $n$-sorters as basic blocks for sorting.

\begin{figure}[!t]
\centering
\includegraphics[width=7.5cm]{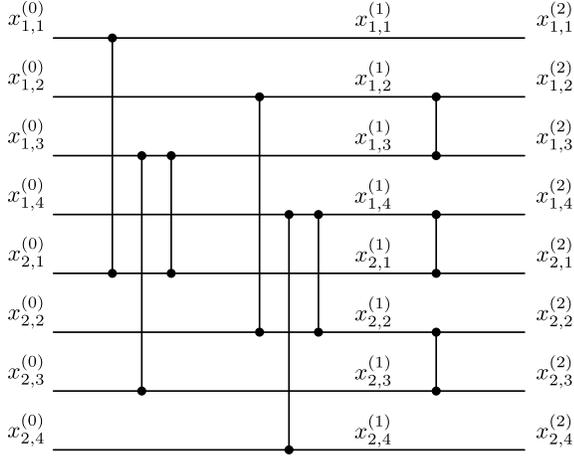}
\caption{The odd-even merge of two sorted lists of 4 values each using 2-sorters.}
\label{fig:oddeven}
\end{figure}

Merging-based sorting networks are an important family of sorting networks, where the merging operation is the key. There are two classes of merging algorithms, the odd-even merging \cite{Bat68sorting} and the bitonic merging \cite{batcher1990bitonic}. The former is an efficient sorting technique based on the divide-and-conquer approach, which decomposes the inputs into two sublists (odd and even), sorts each sublist, and then merges two sorted lists into one. Further decomposition and merging operations are applied on the sublists.
An example of odd-even merging network using 2-sorters is shown in Fig.~\ref{fig:oddeven}, where two sorted lists, $\langle x^{(0)}_{1,1}, \cdots,x^{(0)}_{1,4} \rangle$ and $\langle x^{(0)}_{2,1}, \cdots, x^{(0)}_{2,4} \rangle$, are merged as a single list $\langle x^{(2)}_{1,1}, \cdots,x^{(2)}_{1,4}, x^{(2)}_{2,1}, \cdots, x^{(2)}_{2,4} \rangle$ in two stages.

\begin{figure}[!t]
\centering
\includegraphics[width=8.5cm]{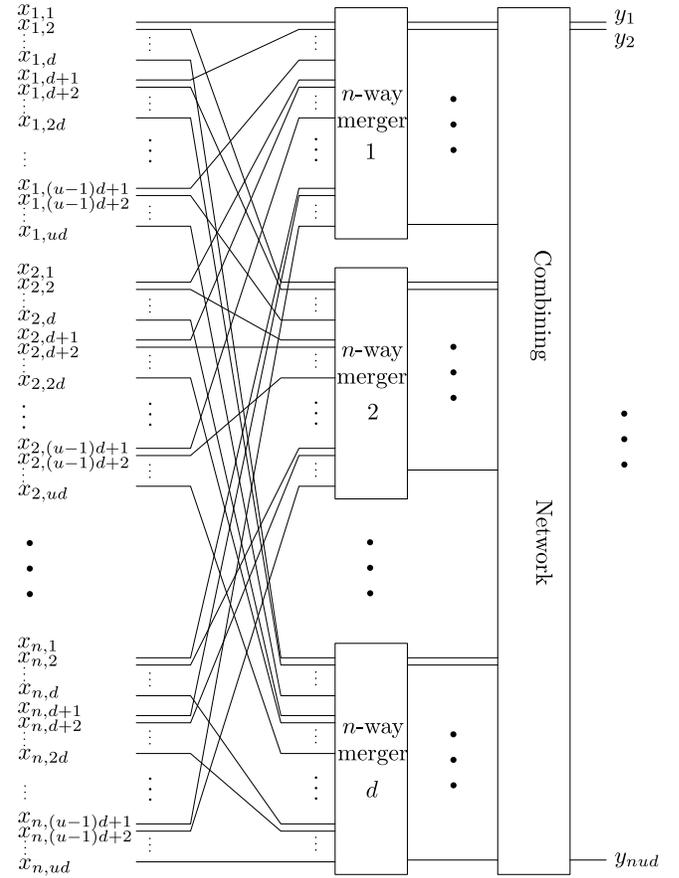}
\caption{Iterative construction rule for the $n$-way merger \cite{Lee95multiway}.}
\label{fig:kway}
\end{figure}

Instead of merging two lists, multiple sorted lists can be merged as a single sorted list simultaneously.
An \textbf{$n$-way merger} ($n\ge 2$) of size $m$ is a network merging $n$ sorted lists of size $m$ ($m$ values) each into a single sorted list in multiple stages.
This was first proposed as a generalization of the Batcher's odd-even merging algorithm. It is also motivated by the use of $n$-sorters, which sort $n$ ($n\ge 2$) values in unit time \cite{parker1989constructing,beigel1990sorting}. Since large sorters are used as basic building blocks, the number of sorters as well as the latency is expected to be reduced greatly.
Many multiway merging algorithms exist in the literature \cite{drysdale1975improved, van1974economical, Lee95multiway, nakatani1989e1, lee1994sorting, leighton1984tight, parker1989constructing, liszka1993generalized, gao1997sloping, zhao1998efficient}. The algorithms in \cite{drysdale1975improved, van1974economical} implement multiway merge using 2-sorters. In \cite{Lee95multiway}, a generalization of Batcher's odd-even is introduced as shown in Fig.~\ref{fig:kway}, where an $n$-way merger of $n$ lists of size $ud$ is decomposed into $d$ $n$-way mergers of $n$ sublists of size $u$ plus a combining network. Each of the small $n$-way mergers is further decomposed similarly. However, the combining network in the merging network in Fig.~\ref{fig:kway} still uses 2-sorters as basic blocks.
In \cite{leighton1984tight}, Leighton proposed a columnsort algorithm, which showed how to sort an $m \times n$ matrix denoting the $n$ sorted lists of $m$ values each. A modification of Leighton's columnsort algorithm was given in \cite{parker1989constructing}.
In \cite{gao1997sloping,zhao1998efficient}, merging networks with $n$-sorters as basic blocks are introduced based on the modified Leighton's columnsort algorithm.

In this work, we focus on multiway merge sort with binary values as inputs. Our merge sort also works for arbitrary values, which is justified by the following theorem.
\begin{thm}[Zero-one principle~\cite{Bat68sorting}]
  If a network with $n$ input lines sorts all $2^n$ lists of 0s and 1s into nondecreasing order, it will sort any arbitrary list of $n$ values into nondecreasing order.
\end{thm}

\section{Multiway Merging}
\label{sec:kmerge}
In the following, we propose an $n$-way merging algorithm with $n$-sorters as basic building blocks as shown in Alg.~\ref{alg:nway}.
We consider a sorting network, where all iterations of Alg.~\ref{alg:nway} are simultaneously instantiated (loop unrolling). We refer to the instantiation of iteration $i$ of Alg.~\ref{alg:nway} as stage $i$ of the sorting network.
The sorters in the last for loop in Alg.~\ref{alg:nway} consist of the last stage.
Let the $n$ sorted input lists be $\langle x^{(0)}_{j,1}, x^{(0)}_{j,2}, \cdots x^{(0)}_{j,m} \rangle$ for $j=1,\cdots,n$. Denote the values of $j$-th list after stage $k$ by $(x^{(k)}_{j,1}, x^{(k)}_{j,2},\cdots, x^{(k)}_{j,m})$. After $T=1+\lceil \frac{m}{2} \rceil$ stages, all input lists are sorted as a single list, $\langle x^{(T)}_{1,1}, x^{(T)}_{1,2},\cdots, x^{(T)}_{1,m} \rangle, \langle x^{(T)}_{2,1}, x^{(T)}_{2,2},\cdots, x^{(T)}_{2,m} \rangle, \cdots, \langle x^{(T)}_{n,1}$, $x^{(T)}_{n,2},\cdots, x^{(T)}_{n,m} \rangle$.

For convenience of describing and proving our algorithm, we introduce some notations and definitions. Denote the number of zeros in the $j$-th list after stage $i$ as $r^{(i)}_j$, where $i=1,2,\cdots, \lceil \frac{m}{2}\rceil + 1$ and $j=1,\cdots,n$. A sorter is called a \textbf{$k$-spaced sorter} if its adjacent inputs span $k$ other wires and each connection of the same sorter comes from different lists of $m$ wires, where $0 \le k \le m-1$. For simplicity, we arrange the sorters in the order of their first connections in each stage. Denote $\{1,2,\cdots,m\}$ as $\mathbb{Z}_m$. Two $k$-spaced sorters are said to be \textbf{adjacent} if they connect adjacent two wires, $x^{(i)}_{j,k}$ and $x^{(i)}_{j,k+1}$, respectively, for some $j \in \mathbb{Z}_m$ and $k \in \mathbb{Z}_{m-1}$.
Then, our $n$-way merging Alg.~\ref{alg:nway} can be intuitively understood as flooding lists with zeros in descending order. The correctness of Alg.~\ref{alg:nway} can be shown by first proving the following lemmas. See the appendix for the proofs of the following lemmas and theorems.

\begin{algorithm}[!tp]
  \caption{Algorithm for $n$-way merging network.}
  \begin{algorithmic}
    \REQUIRE $n$ sorted lists $\langle x^{(0)}_{j,1}, x^{(0)}_{j,2}, \cdots x^{(0)}_{j,m} \rangle$ for $j=1,\cdots,n$;
    \STATE $i=1$;
    \WHILE{$i \le \lceil \frac{m}{2} \rceil$}
        \FOR{$j=1$ to $n-1$}
            \STATE Apply $(m-i)$-spaced sorters between lists $j$ and $j+1$;
        \ENDFOR
        \STATE Merge all $(m-i)$-spaced sorters;
        \STATE Update $n$ sorted lists $\langle x^{(i)}_{j,1}, x^{(i)}_{j,2}, \cdots x^{(i)}_{j,m} \rangle$ for $j = 1, \cdots, n$;
        \STATE $i = i+1$;
    \ENDWHILE
    \FOR{$j=1$ to $n-1$}
        \STATE Apply $(m-1)$-sorters on $m-1$ adjacent lines with first half, $x^{(i-1)}_{j,m-k}$, from list $j$ and second half, $x^{(i-1)}_{j+1,k}$, from list $j+1$, where $k=1,\cdots,\frac{m-1}{2}$;
    \ENDFOR
    \RETURN Sorted lists.
  \end{algorithmic}
  \label{alg:nway}
\end{algorithm}

\begin{figure}[!t]
\centering
\includegraphics[width=8.5cm]{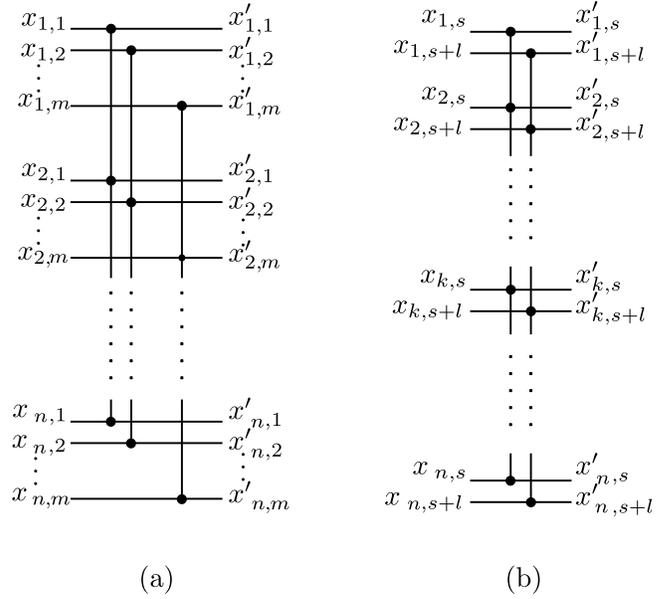}
\caption{The network for $n$ sorted lists of $m$ wires.}
\label{fig:mnsort}
\end{figure}

\begin{lemma}
  Apply $(m-1)$-spaced sorters to $n$ lists of $m$ values, $\langle x_{j,1}, x_{j,2}, \cdots, x_{j,m} \rangle$, for $j=1,\cdots,n$. The outputs of each list are still sorted, $\langle  x'_{j,1}, x'_{j,2}, \cdots, x'_{j,m} \rangle$, for $j=1,2,\cdots,n$.
\label{lm:subgroupsort}
\end{lemma}
For $n$ sorted lists of $m$ values, there are $m$ $(m-1)$-spaced sorters as illustrated in Fig.~\ref{fig:mnsort}(a). The proof of the lemma can be reduced to showing that any two wires $s, s+l \in \mathbb{Z}_m$ of each list connected by the $s$- and $(s+l)$-th sorters are sorted. The simplified network is shown in Fig.~\ref{fig:mnsort}(b). Without lose of generality, we can choose $l=1$.

\begin{figure*}[!t]
\centering
\includegraphics[width=17.5cm]{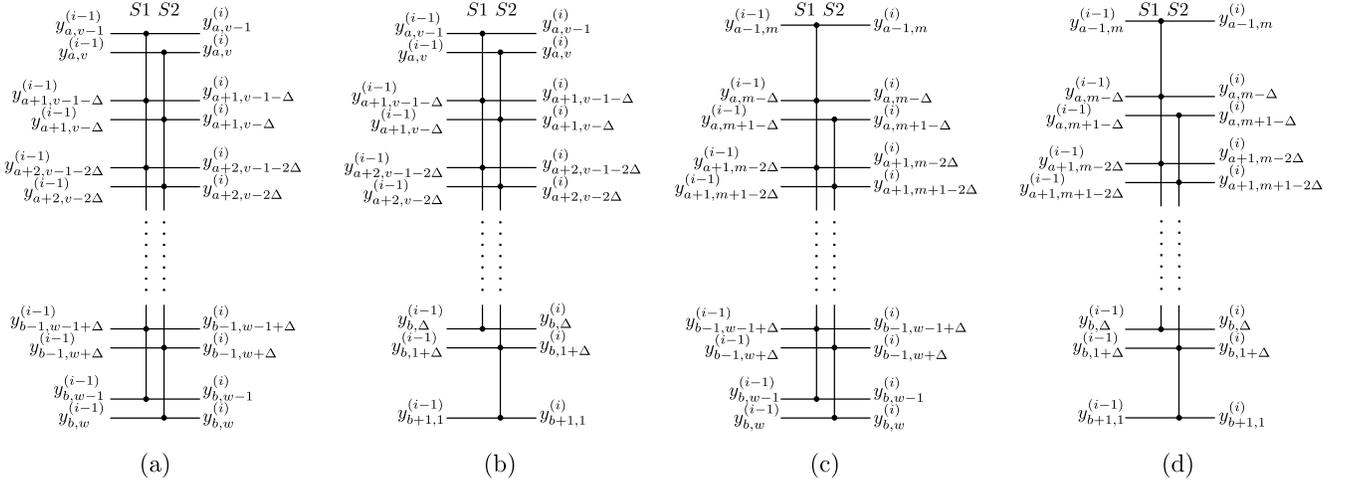}
\caption{Adjacent two sorters $S1$ and $S2$ in each stage of Alg.~\ref{alg:nway} can be classified into four four cases. (a) Case I ($\Delta = \frac{v-w}{b-a}$); (b) Case II ($\Delta = \frac{v-1}{b-a+1}$); (c) Case III ($\Delta = \frac{m-w+1}{b-a+1}$); (d) Case IV ($\Delta = \frac{m}{b-a+2}$).}
\label{fig:m2sort}
\end{figure*}

\begin{lemma}
  In each stage of Alg.~\ref{alg:nway}, there are at most four cases of adjacent two sorters as shown in Fig.~\ref{fig:m2sort}. If $m$ is prime, case IV is impossible.
\label{lm:sorter4case}
\end{lemma}
We first show that the first connections of adjacent two sorters, $S1$ and $S2$, belong to either the same list or adjacent two lists. The same relation is true for the last connections of $S1$ and $S2$. This gives us a total of four cases as shown in Fig.~\ref{fig:m2sort}, where $b\ge a+1$ for Fig.~\ref{fig:m2sort}(a)-(c), and $b\ge a$ for Fig.~\ref{fig:m2sort}(d) such that $S1$ and $S2$ have a size of at least two.

The following theorem proves the correctness of Alg.~\ref{alg:nway}.
\begin{thm}
  For a prime $m$ in Alg.~\ref{alg:nway}, all lists are self-sorted after every stage. In particular, all lists are sorted after the final stage.
\label{thm:groupsort}
\end{thm}
The theorem can be proved by induction on $i$.



In Alg.~\ref{alg:nway}, the latency increases linearly with $\lceil \frac{m}{2}\rceil$. When $m$ is large, the latency is also very large. By further decomposing $m$ into a product of small factors, we can reduce the latency significantly.
In the following, we propose Alg.~\ref{alg:nwaymerge} for merging $n$ lists of $m$ values, where $m=n^{p-1}$ for $p \ge 2$. When $m$ is not a power of $n$, we can use a larger network of $m' = n^{p'} > m$ inputs.
For any $q$ in stage $i$ ($2\le i \le p-1$), denote the number of zeros in each new formed list after stage $i$ as $r^{(i)}_{j,q}$, where $j=1,\cdots,n^i$. Assume two dummy lists with $r^{(i)}_{0,q} = n$ and $r^{(i)}_{n^i+1,q} = 0$ are appended to the two ends of $n^i$ lists.
The correctness of Alg.~\ref{alg:nwaymerge} can be shown by first proving the following lemma.

\begin{algorithm}[!tp]
  \caption{Algorithm for combining $n$ lists of $m=n^{p-1}$ values.}
  \begin{algorithmic}
    \REQUIRE $n$ sorted lists $\langle x^{(0)}_{j,1}, x^{(0)}_{j,2}, \cdots x^{(0)}_{j,m} \rangle$ for $j=1,\cdots,n$ and $m=n^{p-1}$;
    \STATE $i=1$;
        \FOR{$q=1$ to $n^{p-2}$}
            \STATE Apply Alg.~\ref{alg:nway} on $\langle x^{(0)}_{j,q}, x^{(0)}_{j,n^{p-2} + q}, x^{(0)}_{j,2n^{p-2} + q}, \cdots,$ $x^{(0)}_{j,(n - 1) n^{p-2} + q} \rangle$ for $j=1,\cdots,n$ and obtain a single sorted list $\langle x^{(1)}_{1,q}, x^{(1)}_{1,n^{p-2} + q}, \cdots x^{(1)}_{1,(n - 1) n^{p-2} + q}, x^{(1)}_{2,q}, x^{(1)}_{2,n^{p-2} + q}, \cdots$,
            $x^{(1)}_{2,(n - 1) n^{p-2} + q}, \cdots, x^{(1)}_{n,q}$, $x^{(1)}_{n,n^{p-2} + q}, \cdots, x^{(1)}_{n,(n - 1) n^{p-2} + q} \rangle$;
        \ENDFOR
    \FOR{$i=2$ to $p-1$}
        \FOR{$q=1$ to $n^{p-1-i}$}
            \STATE Group $n$ neighboring values of $\langle x^{(i-1)}_{j,q}, x^{(i-1)}_{j,n^{p-i-1} + q}, x^{(i-1)}_{j,2n^{p-i-1} + q}, \cdots x^{(i-1)}_{j,(n - 1) n^{p-i-1} + q} \rangle$ for $j=1,\cdots,n$ and denote the new lists as $\langle x^{(i-1)}_{j,q}, x^{(i-1)}_{j,n^{p-i-1} + q}, \cdots x^{(i-1)}_{j,(n-1)n^{p-i-1} + q} \rangle$ for $j=1,\cdots,n^{i}$;
            \FOR{$k=2$ to $\lceil \frac{n}{2} \rceil$}
                \STATE Apply $(n-k)$-spaced sorters between lists $j$ and $j+1$;
            \ENDFOR
            \STATE Apply $(n-1)$-sorters between lists $j$ and $j+1$ for $j=1,\cdots,n^{i}-1$;
            \STATE Obtain a single sorted list $\langle x^{(i)}_{1,q}, x^{(i)}_{1,n^{p-i-1} + q}, \cdots, x^{(i)}_{1,(n-1)n^{p-i-1} + q}, x^{(i)}_{2,q},$ $x^{(i)}_{2,n^{p-i-1} + q}, \cdots, x^{(i)}_{2,(n-1)n^{p-i-1} + q}, \cdots, x^{(i)}_{n^i,q},$ $x^{(i)}_{n^i,n^{p-i-1} + q}, \cdots, x^{(i)}_{n^i,(n-1)n^{p-i-1} + q}\rangle$;
        \ENDFOR
    \ENDFOR
    \RETURN Sorted list.
  \end{algorithmic}
  \label{alg:nwaymerge}
\end{algorithm}

\begin{lemma}
  In Alg.~\ref{alg:nwaymerge}, the new lists in stage $i$ with respect to $q$ are self-sorted. The numbers of zeros of all new lists after stage $i$ are non-increasing,
  \[
  r^{(i)}_{j,q} \ge r^{(i)}_{j+1,q} \quad \mbox{for} \quad j=1,\cdots, n^i-1,
  \]
  where $i=2,\cdots, p-1$ and $q=1,\cdots,n^{p-1-i}$. Furthermore, there are at most $n$ consecutive lists that have between 1 and $n-1$ zeros,
  \[
    r^{(i)}_{s,q} = n > r^{(i)}_{s+1,q} \ge \cdots \ge r^{(i)}_{s+l,q} > 0 = r^{(i)}_{s+l+1,q} \quad \mbox{for} \quad l \le n,
  \]
  where $s \ge 0$ and $s+l \le n^i$.
\label{lm:mergen0s}
\end{lemma}
See Sec.~\ref{pf:mergen0s} for the proof.

The following theorem proves the correctness of Alg.~\ref{alg:nwaymerge}.
\begin{thm}
  Alg.~\ref{alg:nwaymerge} combines $n$ sorted lists of $m=n^{p-1}$ values as a single sorted list.
  \label{thm:nwaymerge}
\end{thm}
In Alg.~\ref{alg:nwaymerge}, the latency is reduced to $1+(p-1)\lceil \frac{n}{2}\rceil$ for $n$ sorted lists of $m=n^{p-1}$ values.

\begin{figure}[!t]
\centering
\includegraphics[width=8.5cm]{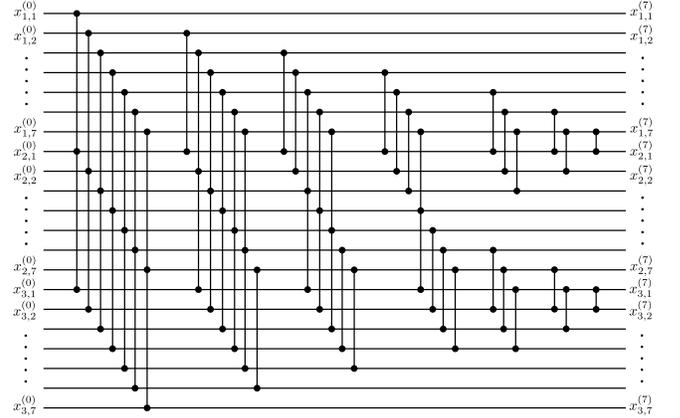}
\caption{A 3-way merging network of $N=3\times 7$ inputs implemented via 7 stages.}
\label{fig:3by7merger}
\end{figure}

\begin{figure}[!t]
\centering
\includegraphics[width=6cm]{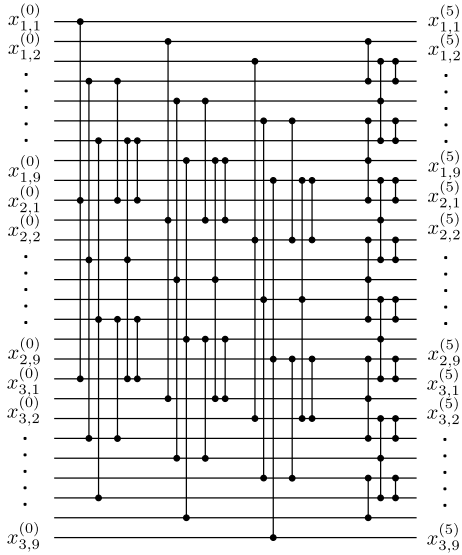}
\caption{A 3-way merging network of $N=3\times 9$ inputs implemented via 5 stages.}
\label{fig:3by9merger}
\end{figure}

In the following, we show two examples for comparison of the two algorithms.
First, a 3-way merging network of $N=3\times 7$ inputs via Alg.~\ref{alg:nway} is shown in Fig.~\ref{fig:3by7merger}.
Then, a 3-way merging network of $N=3\times 9$ inputs via Alg.~\ref{alg:nwaymerge} is shown in Fig.~\ref{fig:3by9merger}. Though there are more inputs in Fig.~\ref{fig:3by9merger} than that in Fig.~\ref{fig:3by7merger}, the latency of Alg.~\ref{alg:nwaymerge} is smaller due to recursive decomposition. The numbers of sorters in Figs.~\ref{fig:3by7merger} and \ref{fig:3by9merger} are given by 40 and 41, respectively. For six more inputs, it requires only one more sorter in Fig.~\ref{fig:3by9merger}. Hence, Alg.~\ref{alg:nwaymerge} can be more efficient than Alg.~\ref{alg:nway} for a large $m$.

\begin{figure}[!t]
\centering
\includegraphics[width=8.5cm]{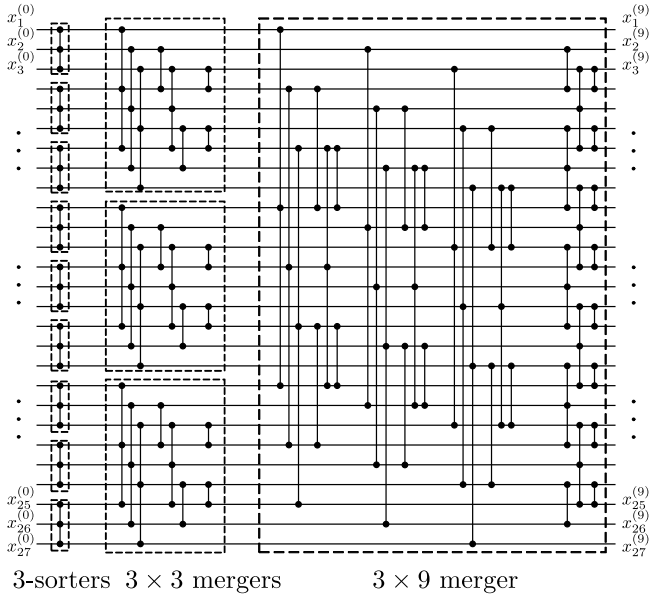}
\caption{A 3-way sorting network of $N=3^3$ inputs implemented via 9 stages.}
\label{fig:27sorter}
\end{figure}

\section{Multiway Sorting}
\label{sec:ksort}
In this section, we first focus on how to construct sorting networks with $n$-sorters using the multiway merging algorithm in Sec.~\ref{sec:kmerge}. Then, we analyze the latency and the number of sorters of the proposed sorting networks by deriving the closed-form expressions. We compare them with previously proposed SS-Mk in \cite{gao1997sloping} but not the ISS-Mk \cite{zhao1998efficient}, because for our interested ranges of $N$, the ISS-Mk requires larger latency due to a large constant.
\subsection{Multiway sorting algorithm}
Based on the multiway merging algorithm in Sec.~\ref{sec:kmerge}, we proposed a parallel sorting algorithm using a divide-and-conquer method. The idea is to first decompose large list of inputs into smaller sublists, then sort each sublist, and finally merge them into one sorted list. The sorting of each sublist is done by further decomposition.
For instance, for $N=n^p$ inputs, we first divide the $n^p$ inputs into $n$ lists of $n^{p-1}$ values. Then we sort each of these $n$ lists and combine them with Alg.~\ref{alg:nwaymerge}. The sorting operation of each of the $n$ lists is done by dividing the $n^{p-1}$ inputs into $n$ smaller lists of $n^{p-2}$ values. We repeat the above operations until that each of $n$ smaller lists contains only $n$ values, which can be sorted by a single $n$-sorter.
The detailed procedures are shown in Alg.~\ref{alg:nwaysort}.

\begin{algorithm}[!h]
  \caption{Algorithm for sorting $N=n^{p}$ values.}
  \begin{algorithmic}
    \REQUIRE $N=n^{p}$ values, $x^{(0)}_{1}, x^{(0)}_{2}, \cdots, x^{(0)}_{n^p}$;
    \STATE Partition the $N=n^p$ values as $n^{p-1}$ lists of $n$ values each, $(x^{(0)}_{j,1}, x^{(0)}_{j,2}, \cdots, x^{(0)}_{j,n})$ for $j=1,\cdots,n^{p-1}$;
    \STATE Apply one $n$-sorter on each of $n^{p-1}$ lists and obtain $\langle x^{(1)}_{j,1}, x^{(1)}_{j,2}, \cdots, x^{(1)}_{j,n} \rangle$ for $j=1,\cdots,n^{p-1}$;
    \FOR{$i = 2$ to $p$}
        \FOR{$j=1$ to $n^{p-i}$}
            \STATE Apply Alg.~\ref{alg:nway} on $\langle x^{(i-1)}_{(j-1)n+k,1}, x^{(i-1)}_{(j-1)n+k,2}, \cdots,$ $x^{(i-1)}_{(j-1)n+k,n^{i-1}} \rangle$ for $k=1,\cdots,n$, and obtain a single sorted list $\langle x^{(i)}_{j,1}, x^{(i)}_{j,2}, \cdots x^{(i)}_{j,n^i} \rangle$;
        \ENDFOR
    \ENDFOR
    \RETURN Sorted list.
  \end{algorithmic}
  \label{alg:nwaysort}
\end{algorithm}

For example, a 3-way sorting network of $N=3^3$ inputs is shown in Fig.~\ref{fig:27sorter}. The first stage contains 9 3-sorters. The second stage contains 3 three-way mergers with a depth of 3. The last stage contains a three-way merger with a depth of 5. The total depth is given by 9.

\subsection{Latency analysis}
First, we focus on the latency for sorting $N$ values. The latency is defined as the number of basic sorters in the longest paths from the inputs to the sorted output. In Alg.~\ref{alg:nwaysort}, there are $p$ iterations. In iteration $i$, there are $n^i$ merging networks, each of which is to merge $n$ sorted lists of $n^{p-i}$ values. For iteration $i$, the latency is given by $L_{our}(n,n^{i-1}) = 1+(i-1)\lceil \frac{n}{2} \rceil$.
For a sorting network of $N=n^p$ values via Alg.~\ref{alg:nwaysort}, by summing up the latencies of all levels, we obtain the total latency
\begin{equation}
\begin{array}{rcl}
L_{our}(n^p) &=& \sum^{p}_{i=1} L_{our}(n,n^{i-1}) \\
&=& p + \lceil \frac{n}{2} \rceil \times \frac{p(p-1)}{2}.
\end{array}
\label{eq:our_L}
\end{equation}
The closed-form expression of latency for the SS-Mk given in \cite{gao1997sloping} is
\begin{equation}
L_{SS-Mk}(n^p) = 1+(p-1)n + \frac{(p-1)(p-2)}{2}\lceil \log_2 n \rceil.
\label{eq:ssMk_L}
\end{equation}

We compare our latency for sorting $N=n^p$ values with that for the SS-Mk in \cite{gao1997sloping}. From Eqs.~(\ref{eq:our_L}) and (\ref{eq:ssMk_L}), for $N=n^p$ inputs, $p$ should be as small as possible to obtain small latencies.
In Table~\ref{tab:latency}, we compare the latencies of Eqs.~(\ref{eq:our_L}) and (\ref{eq:ssMk_L}) for small $p$ ($p=2,3,4$). It is easily seen that our implementation has a smaller latency than the SS-Mk in \cite{gao1997sloping} for a prime greater than 3. It is also observed that $L_{our}(2^p) = L_{SS-Mk}(2^p) = p(p+1)/2$ for $n=2$, which is the same as the odd-even merge sort in \cite{Bat68sorting}.

\begin{table}[!th]
\caption{Comparison of latencies of sorting networks of $N=n^p$ inputs via the SS-Mk in \cite{gao1997sloping} and our implementation.}\label{tab:latency}
\begin{center}
\begin{tabular}{|c|c|c|c|}
\hline
& $p=2$ & $p=3$ & $p=4$ \\
\hline
\cite{gao1997sloping} & $1+n$ & $1+2n + \lceil \log_2 n \rceil$ & $1+3n + 3 \lceil \log_2 n \rceil$ \\
\hline
Ours & $2+\lceil \frac{n}{2} \rceil$ & $3 + 3\lceil \frac{n}{2} \rceil$ & $4+ 6\lceil \frac{n}{2} \rceil$\\
\hline
\end{tabular}
\end{center}
\end{table}

\subsection{Analysis of the number of sorters}
\label{sec:compsorter}
In the following, we compare the number of sorters of our algorithms with the SS-Mk in \cite{gao1997sloping}.
Since the distribution of sorters for an arbitrary sorting network of $N$ inputs is not known, we assume that any $m$-sorter ($m < n$) has the same delay and area as the basic $n$-sorter and count the number of sorters.
We first derive the closed-form expression of the number of sorters for sorting $N$ values via our Alg.~\ref{alg:nwaysort}.
Since the expression of the number of sorters for the SS-Mk was not provided in \cite{gao1997sloping}, we also derive the corresponding closed-form expression and compare it with our algorithm. The whole sorting network is constructed recursively by merging small sorted lists into a larger sorted list. We first derive the number of sorters of a merging network of $n$ lists of $n^{p-i}$ values, which is given by
\[
S_{our}(n,n^{p-i}) = (p-i)\cdot M^*_{n^{p-i}} + \frac{n^{p-i}-1}{n-1}\cdot C^*_n + n^{p-i},
\]
where $M^*_{n^{p-i}} = \left( 1+ \frac{\lceil n/2 \rceil (\lceil n/2 \rceil -1) }{2} \right)n^{p-i}$ and $C^*_n = (\lceil n/2 \rceil -1)n - \frac{3\lceil n/2 \rceil (\lceil n/2 \rceil -1)}{2} -1$.
By summing up the numbers of sorters of all mergers in all stages, we obtain the total number of sorters, which is given by
\begin{equation}
\begin{array}{rcl}
T_{our}(n^p) &=& \sum^{p-1}_{i=1} n^{i-1} \cdot S_{our}(n,n^{p-i}) + n^{p-1}\\
&=& \frac{p(p-1)}{2}\cdot M^*_{n^{p-1}} + \left[ \frac{(p-1)n^{p-1}}{n-1} - \frac{n^{p-1}-1}{(n-1)^2} \right] \\
& & \cdot C^*_n + pn^{p-1},
\end{array}
\label{eq:our_T}
\end{equation}
As $N \rightarrow \infty$, $T_{our}(n^p)$ is on the order of $O(A_1 \frac{N\log N (\log N - \log n)}{(\log n)^2 /n} + A_2 \frac{N (\log N -\log n)}{\log n} + A_3 \frac{N \log N}{n \log n})$.
Similarly for the SS-Mk in \cite{gao1997sloping}, the number of sorters of the merging network of $n$ lists of $n^{p-i}$ values each is given by
\[
S_{SS-Mk}(n,n^{p-i}) = M^\dagger_{n^{p-i}} + K^\dagger_{n,n^{p-i}} + C^\dagger_n,
\]
where $M^\dagger_{n^{p-i}} = \Big( \frac{(n+1-\lceil n/2 \rceil)(n-\lceil n/2 \rceil)}{2} + \frac{(\lceil n/2 \rceil+1)(\lceil n/2 \rceil-2)}{2}$ $ + 2 \Big)n^{p-i}$, $K^\dagger_{n,n^{p-i}} = \lceil \log_2 n^{p-1-i} \rceil n^{p-i} + (n-3)2^{\lceil \log_2 n^{p-1-i} \rceil +1}$, and $C^\dagger_n = (\lceil n/2 \rceil -2)n - \frac{3(\lceil n/2 +1 \rceil) (\lceil n/2 \rceil -2)}{2} - \frac{(n+1-\lceil n/2 \rceil)(n-\lceil n/2 \rceil)}{2} - (n-3)$.
The total number of sorters of the sorting network via the SS-Mk in \cite{gao1997sloping} is given by
\begin{equation}
\begin{array}{rcl}
T_{SS-Mk}(n^p) &=& \sum^{p-1}_{i=1} n^{i-1} \cdot S_{SS-Mk}(n,n^{p-i}) + n^{p-1}\\
&=& (p-1)\cdot M^\dagger_{n^{p-1}} + \frac{n^{p-1} - 1}{n-1} \cdot C^\dagger_n  + n^{p-1} \\
& & + n^{p-1}\sum^{p-2}_{i=1}\lceil i \log_2 n \rceil \\
& & + \sum^{p-1}_{i=1} n^{i-1}(n-3)2^{\lceil (p-1-i) \log_2 n \rceil + 1},
\end{array}
\label{eq:ssMk_T}
\end{equation}
As $N \rightarrow \infty$, $T_{SS-Mk}(n^p)$ is on the order of $O(B_1 \frac{N(\log N - \log n)}{(\log n) /n} + B_2 \frac{N \log N (\log N -\log n)}{n \log n} + B_3 \frac{N (\log N -\log n)}{n \log n} + B_4 \frac{N}{n})$.

According to the big-O expressions of $T_{our}(n^p)$ and $T_{SS-Mk}(n^p)$, when $n$ is bounded, the asymptotic bounds on the number of sorters required by both our Alg.~\ref{alg:nwaysort} and the SS-Mk in \cite{gao1997sloping} are given by $O(N \log^2 N)$, which is also the asymptotical bound for the odd-even and bitonic sorting algorithms~\cite{Bat68sorting, batcher1990bitonic}.
When $N$ is fixed and $n$ increases, the first term of the big-O expressions of $T_{our}(n^p)$ and $T_{SS-Mk}(n^p)$ decreases first, then increases, and decreases to zero when $n \rightarrow N$. While other terms decrease monotonically with $n$. Hence, if $n$ is not constrained, the minimum value of $T_{our}(n^p)$ and $T_{SS-Mk}(n^p)$ is one when $n=N$, meaning a single $N$-sorter is used.

\subsection{Comparison of the number of sorters}
\label{sec:comp}
According to the analysis of both our Alg.~\ref{alg:nwaysort} and the SS-Mk in \cite{gao1997sloping}, the number of sorters for sorting $N=n^p$ inputs can be reduced by using a larger basic sorter. However, a very large basic sorter is not feasible due to some practical concerns, such as fan-in and cost. In this work, we assume that the basic sorter size is limited.
For a given $N$, we take the total number of sorters in Eqs.~(\ref{eq:our_T}) and (\ref{eq:ssMk_T}) as a function of $p$ with $n=N^{1/p} \le n_b$, where $n_b$ is the upper bound of the basic sorter size.
When $N$ is not a power of a prime, we append redundant inputs of 0's and get a larger $N'$ such that $N'$ is a power of a prime. Hence, we have $n'= N'^{1/p} = \lceil\lceil N^{1/p} \rceil \rceil$, where $\lceil \lceil x \rceil \rceil$ denotes the smallest prime larger than or equal to $x$.
There exists an optimal $p$ such that the total number of sorters is the minimum.
We search for the optimal $p$'s for our Alg.~\ref{alg:nwaysort} and the SS-Mk~\cite{gao1997sloping} using MATLAB. By plugging the optimal $p$'s into Eqs.~(\ref{eq:our_T}) and (\ref{eq:ssMk_T}), we obtain the total number of sorters for sorting networks of $N$ inputs.

\begin{figure}[!h]
\centering
\includegraphics[width=8.5cm]{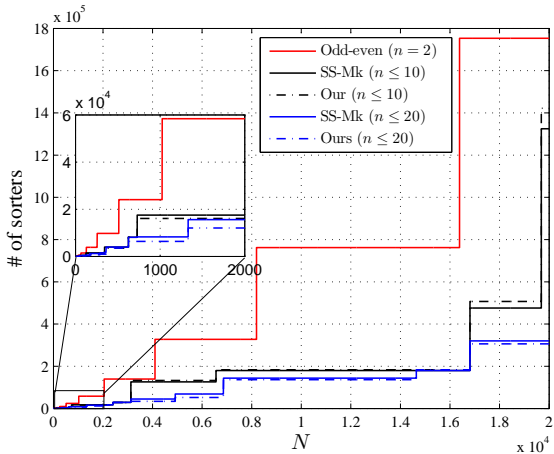}
\caption{Comparison of the number of sorters ($n\le 10$ and $n\le 20$) for sorting $N$ inputs via the SS-Mk in \cite{gao1997sloping} and our Alg.~\ref{alg:nwaysort}.}
\label{fig:complexity_n1020}
\end{figure}

We compare the number of sorters for sorting networks via the Batcher's odd-even algorithm \cite{Bat68sorting}, our Alg.~\ref{alg:nwaysort}, and the SS-Mk \cite{gao1997sloping} for wide ranges of $N$. The results are show in Fig.~\ref{fig:complexity_n1020}. The numbers of sorters are illustrated by staircase curves, because we use a larger sorting network for $N$ not being a power of prime.
From Fig.~\ref{fig:complexity_n1020}, the Batcher's odd-even algorithm using 2-sorters always requires more sorters than both our Alg.~\ref{alg:nwaysort} and the SS-Mk in \cite{gao1997sloping}.
For both our Alg.~\ref{alg:nwaysort} and the SS-Mk \cite{gao1997sloping}, the number of sorters is smaller for a larger $n_b$, meaning that using larger basic sorters reduces the number of sorters.
For the comparison of the number of sorters required by our Alg.~\ref{alg:nwaysort} and the SS-Mk \cite{gao1997sloping}, there are three scenarios with respect to three ranges of $N$. We first focus on $n_b = 10$. For $N \le 6.25 \times 10^2$, our Alg.~\ref{alg:nwaysort} has fewer or the same number of sorters than the SS-Mk as shown in Fig.~\ref{fig:complexity_n1020}. For some segments in $6.25 \times 10^2 < N \le 3.13 \times 10^3$, our Alg.~\ref{alg:nwaysort} has fewer sorters than the SS-Mk.
For $N > 3.13 \times 10^3$, the SS-Mk in \cite{gao1997sloping} needs fewer sorters.
For $n_b = 20$, we have similar results.
For $N \le 1.46 \times 10^4$, our Alg.~\ref{alg:nwaysort} has fewer or the same number of sorters than the SS-Mk as shown in Fig.~\ref{fig:complexity_n1020}. For some segments in $1.46 \times 10^4 < N < 1.3 \times 10^5$, our Alg.~\ref{alg:nwaysort} has fewer sorters than the SS-Mk. For $N > 1.3 \times 10^5$, the SS-Mk in \cite{gao1997sloping} needs fewer sorters.

\begin{figure}[!h]
\centering
\includegraphics[width=8.5cm]{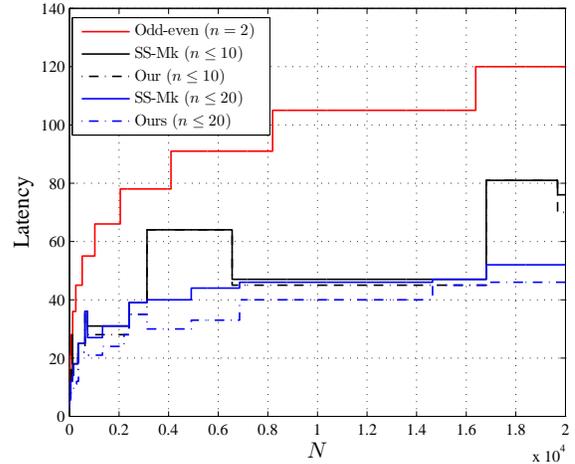}
\caption{Comparison of the latency for sorting $N$ inputs with $n\le 10$ and $n\le 20$ via the SS-Mk in \cite{gao1997sloping} and our Alg.~\ref{alg:nwaysort}.}
\label{fig:latency_n1020}
\end{figure}

Similarly, we compare the latency of the Batcher's odd-even algorithm, our Alg.~\ref{alg:nwaysort}, and the SS-Mk in \cite{gao1997sloping}. The latencies are obtained by plugging the corresponding optimal $p$'s into Eqs.~(\ref{eq:our_L}) and (\ref{eq:ssMk_L}) and shown in Fig.~\ref{fig:latency_n1020} for $N \le 2 \times 10^4$.
From Fig.~\ref{fig:latency_n1020}, the Batcher's odd-even algorithm using 2-sorters has the largest latency. For both our Alg.~\ref{alg:nwaysort} and the SS-Mk \cite{gao1997sloping}, the latency can be reduced by having a larger $n_b$.
The latency of our Alg.~\ref{alg:nwaysort} is not greater than the SS-Mk for $N \le 2 \times 10^4$ for both $n_b = 10$ and $n_b = 20$ as shown in Fig.~\ref{fig:latency_n1020}. This is because our Alg.~\ref{alg:nwaysort} tends to use large sorters, leading to less stages of sorters. We note that the latency goes up and down for some $N$ in Fig.~\ref{fig:latency_n1020}. This is because of the switching from a smaller basic sorter to a larger one to reduce the number of sorters.

To some researchers' interest, we also compare the number of sorters for $N$ being a power of two. The results are shown in Table~\ref{tab:complexity_n20}, where columns two and three show the numbers of sorters for the SS-Mk and our Alg.~\ref{alg:nwaysort}, respectively, and column five shows the reduction by our Alg.~\ref{alg:nwaysort} compared with the SS-Mk \cite{gao1997sloping}. For our Alg.~\ref{alg:nwaysort}, there are up to 46\% fewer sorters than the SS-Mk in \cite{gao1997sloping} for $N = 2^i$, for $i=4,5,\cdots,16$.
It is also observed that a greater reduction is obtained for small $p$, meaning our approach is more efficient for networks with larger sorters as basic blocks.

\begin{table}[!t]
\caption{Comparison of the number of sorters for sorting $N=2^k$ inputs ($1\le k \le 16$) with $n\le 20$ via the SS-Mk in \cite{gao1997sloping} and our Alg.~\ref{alg:nwaysort}.}\label{tab:complexity_n20}
\begin{center}
\begin{tabular}{|c|c|c|c|}
\hline
$N$ & SS-Mk & Ours & Rd. (\%)\\
\hline
2	&	1   &   1	&	0.0	\\
\hline
4	&	5   &   5	&	0.0	\\
\hline
8	&	11   &     11	&	0.0	\\
\hline
16	&	38  &   30	&	21.05\\
\hline
32	&	95  &   65	&	31.58\\
\hline
64	&	347    &   207	&	40.35\\
\hline
128	&	566   &   326	&	42.40\\
\hline
256	&	1250	&    690	&	44.80\\
\hline
512	&	3952   &   3500	&	11.44\\
\hline
1024	&	8287   &   6378	&	23.04	\\
\hline
2048	&	15595	&   12039	&	22.80 \\
\hline
4096	&	44652	&   33891	&	24.10 \\
\hline
8192	&	143762	&   136574	&	5.00\\
\hline
16384   &   179631  &   183143  &   -1.96 \\
\hline
32768   &   1176250 &   1134692 &   3.53 \\
\hline
65536   &   1176250 &   1134692 &   3.53 \\
\hline
\end{tabular}
\end{center}
\end{table}

\section{Application in Threshold Logic}
\label{sec:appl}
In Sec.~\ref{sec:comp}, we assume all basic sorters in the sorting network are the same and measure the complexity by the number of sorters, since the distribution of sorters is unknown. This would overestimate the total complexity.
In this section, we focus on the threshold logic and measure the complexity by the number of threshold gates.
In the following, we first briefly introduce the threshold logic, which is very powerful for computing complex functions, such as parity function, addition, multiplication, and sorting, with significantly reduced number of gates. Then, we present an implementation of a large sorter in threshold logic. Last, we compare the complexity of sorting networks in terms of the number of gates.
This is a very narrow application in the sense that sorters are implemented by threshold logic and the inputs are binary values.

\subsection{Threshold logic}
\begin{figure}[!h]
\centering
\includegraphics[width=3.0cm]{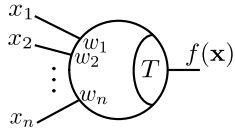}
\caption{Threshold gate realizing $f(\mathbf{x})$ for $n$ inputs, $x_1, x_2, \cdots , x_n$, with corresponding weights $\omega_1, \omega_2, \cdots, \omega_n$ and a threshold $T$.}
\label{fig:TG}
\end{figure}

A threshold function \cite{Mur71} $f$ with $n$ inputs ($n\ge 1$), $x_1, x_2, \cdots, x_n$, is a Boolean function whose output is determined by
\begin{equation}
  f(x_1, x_2, \cdots, x_n) = \left\{ \begin{array}{cl} 1 & \mbox{if } \sum_{i=1}^n  w_ix_i \ge T \\
  0 & \mbox{otherwise,} \end{array} \right.
\label{eq:threshold-def}
\end{equation}
where $w_i$ is called the {\em weight} of $x_i$ and $T$ the {\em threshold}. In this paper we denote this threshold function as $[x_1, x_2, \cdots, x_n; w_1, w_2, \cdots, w_n; T]$, and for simplicity sometimes denote it as $f=[\mathbf{x};\mathbf{w};T]$, where  $\mb{x}=(x_1, x_2, \cdots, x_n)$ and $\mb{w}=(w_1, w_2, \cdots, w_n)$. The physical entity realizing a threshold function is called a threshold gate, which can be realized with CMOS or nano technology. Fig.~\ref{fig:TG} shows the symbol of a threshold gate realizing (\ref{eq:threshold-def}).

\subsection{$n$-sorter}
\begin{figure}[!h]
\centering
\includegraphics[width=7.5cm]{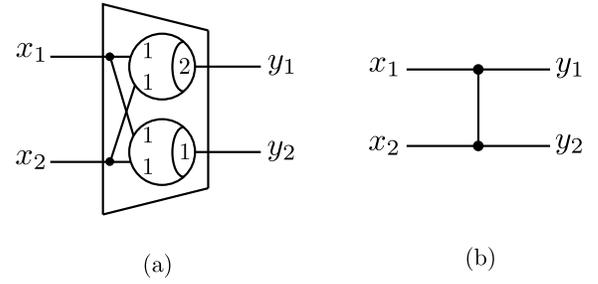}
\caption{Sorters implemented in threshold logic (a) 2-sorter; (b) $n$-sorter.}
\label{fig:sorterTH}
\end{figure}

Binary sorters can be easily implemented in threshold logic. In \cite{Beiu1993enhanced}, a 2-by-2 comparator (2-sorter) was implemented by two threshold gates as shown in Fig.~\ref{fig:sorterTH}(a). Similarly, we introduce a threshold logic implementation of an $n$-sorter as shown in Fig.~\ref{fig:sorterTH}(b), where $n$ threshold gates are required. As shown in Fig.~\ref{fig:sorterTH}, the number of gates of an $n$-sorter scales linearly with the number of inputs $n$.
Hence, large sorters are preferred to be used as basic blocks. However, larger sorters are more complex and expensive to be implemented.
For practical concerns, such as fan-in and cost, some limit on the size of basic sorters is assumed.

\subsection{Analysis of number of gates}
In the following, we assume all gates are the same and derive the total number of gates. The sorting network of $N$ inputs is composed of multiple stages, of which each partially sorts $N$ values. Not all values in each stage participate the comparison-and-switch operation.
A simple way to count the gates is to insert buffer gates in each stage to store values without involving any sorting operation. Buffer insertion is also needed for implementation of threshold logic in some nanotechnology, where synchronization is required for correction operation.
Hence, each stage contains $N$ gates and the total number of gates is obtained by multiplying $N$ to the latency.
Note that $N$ does not have to be a power of $n$.
Hence, the total number of gates of our Alg.~\ref{alg:nwaysort} and the SS-Mk \cite{gao1997sloping} are simply given by
\begin{equation}
Q_{our}(N) = N \cdot L_{our}(N),
\label{eq:Gour_L}
\end{equation}
and
\begin{equation}
Q_{SS-Mk}(N) = N \cdot L_{SS-Mk}(N).
\label{eq:GssMk_L}
\end{equation}
If $n$ is bounded, the total numbers of gates in Eqs.~(\ref{eq:Gour_L}) and (\ref{eq:GssMk_L}) have an order of $O(N \log^2 N)$, which is the same as the order for the numbers of sorters via our Alg.~\ref{alg:nwaysort} and the SS-Mk in \cite{gao1997sloping} in Sec.~\ref{sec:compsorter}.

To derive the accurate number of gates, we first derive the number of buffers added for Eqs.~(\ref{eq:Gour_L}) and (\ref{eq:GssMk_L}).
When $N$ is a power of prime, the number of buffers for sorting $N=n^p$ values via our Alg.~\ref{alg:nwaysort} and the SS-Mk \cite{gao1997sloping} can be easily obtained due to a regular structure.
For our Alg.~\ref{alg:nwaysort}, the number of buffers is given by $G_{our}(N) = (p-1)n^{p-2}\frac{n^2+6n-5}{4} + \frac{((p-2)n^{p-1} - (p-1)n^{p-2} + 1)(n+5)}{4(n-1)} + \frac{(p-1)(p-2)}{2}n^{p-1}$ for $n\neq 2$ and $G(n^p) = (p^2-p+4)2^{p-1}-2$ for $n=2$. Similarly, we derive the number of buffers for the SS-Mk in \cite{gao1997sloping}, which is given by $G_{SS-Mk}(N) = 2\sum^p_{i=2}(2^{\lceil (i-2)\log_2 n \rceil + 1} -1)n^{p-i} + \frac{(n^{p-1}-1)(n^2-5)}{2(n-1)} + \frac{(p-1)(n-1)^2 n^{p-1}}{4}$ for $n\neq 2$ and $G(n^p) = (p^2-p+4)2^{p-1}-2$ for $n=2$.
By subtracting the number of buffers from Eqs.~(\ref{eq:Gour_L}) and (\ref{eq:GssMk_L}), we obtain the total numbers of gates for our algorithm and the SS-Mk as shown in the following,
\begin{equation}
R_{our}(n^p) = n^p \cdot L_{our}(n^p) - G_{our}(n^p),
\label{eq:GBour_L}
\end{equation}
and
\begin{equation}
R_{SS-Mk}(n^p) = n^p \cdot L_{SS-Mk}(n^p) - G_{SS-Mk}(n^p).
\label{eq:GBssMk_L}
\end{equation}
Though it would overestimate the total number of gates by adding buffers. However, the asymptotic gate counts are not affected, since both $G_{our}(n^p)$ and $G_{SS-Mk}(n^p)$ have the same order of $O(N \log^2 N)$.

\subsection{Comparison of the number of gates}
\begin{figure}[!t]
\centering
\includegraphics[width=8.5cm]{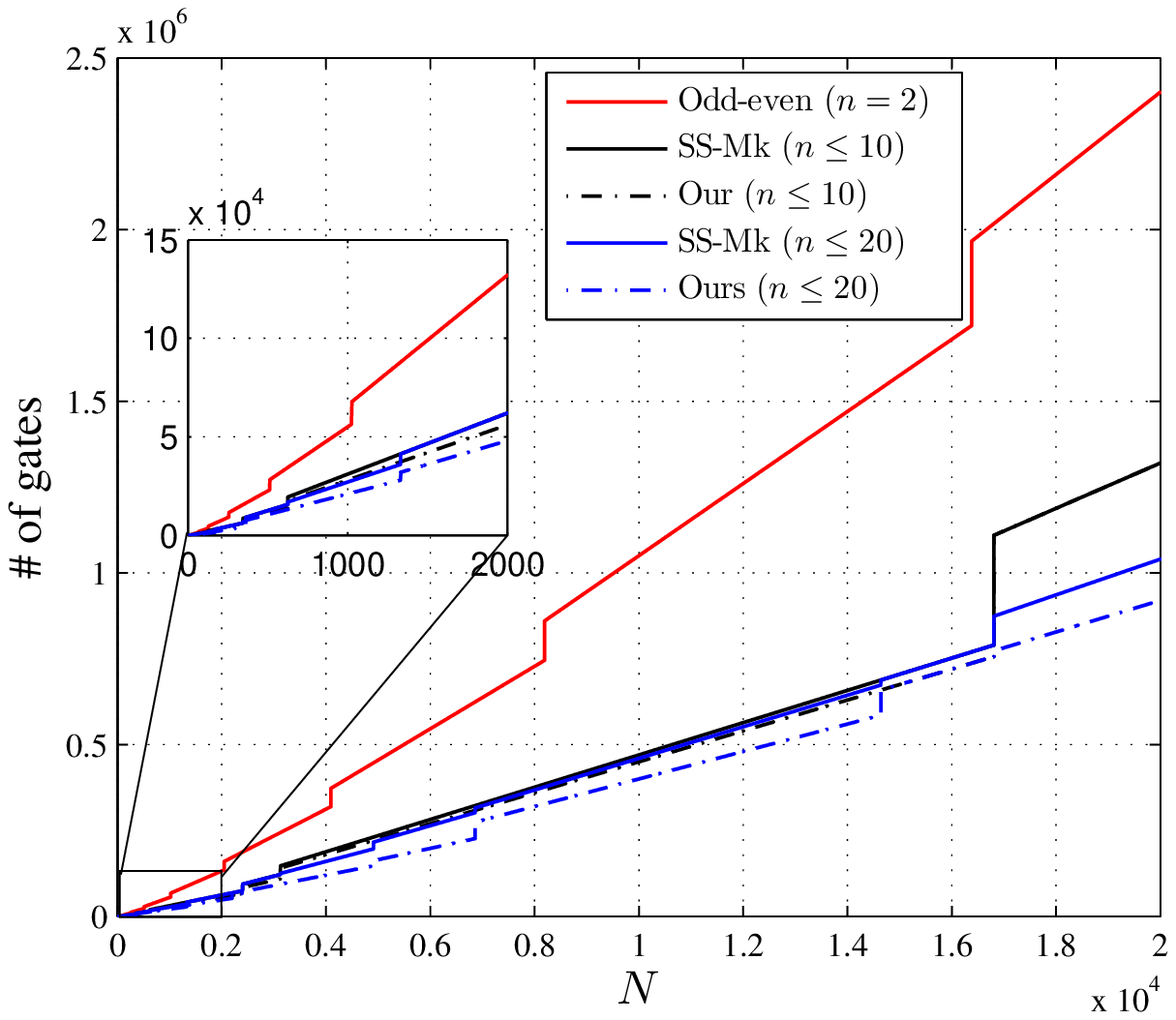}
\caption{Comparison of the number of gates ($n\le 10$ and $n\le 20$) for sorting $N$ inputs via the SS-Mk in \cite{gao1997sloping} and our Alg.~\ref{alg:nwaysort}.}
\label{fig:comp_cap_n1020TG}
\end{figure}

In the following, we first compare the number of gates with consideration of buffers.
Using the same idea as in Sec.~\ref{sec:compsorter}, we search for the optimal $p$'s of Eqs.~(\ref{eq:Gour_L}) and (\ref{eq:GssMk_L}) using MATLAB.
For $n\le 10$ and $n\le 20$, the numbers of gates of the SS-Mk and our two implementations are illustrated in Fig.~\ref{fig:comp_cap_n1020TG}. We also plot the odd-even sorting for comparison.
The curves in Fig.~\ref{fig:comp_cap_n1020TG} are segmented linear lines. This can be explained by Eqs.~(\ref{eq:Gour_L}) and (\ref{eq:GssMk_L}), which are functions of $N$ and latency.
From Fig.~\ref{fig:comp_cap_n1020TG}, the Batcher's odd-even algorithm using 2-sorters has more gates than both our algorithm and the SS-Mk in \cite{gao1997sloping}.
For both our Alg.~\ref{alg:nwaysort} and the SS-Mk \cite{gao1997sloping}, the number of gates  is smaller with a larger $n_b$, meaning that using larger basic sorters reduces the number of gates.
For the comparison of the number of gates required by our Alg.~\ref{alg:nwaysort} and the SS-Mk \cite{gao1997sloping}, there are also three scenarios with respect to three ranges of $N$.
We first focus on $n_b = 10$. For $N \le 1.68 \times 10^4$, our Alg.~\ref{alg:nwaysort} has fewer or the same number of gates than the SS-Mk as shown in Fig.~\ref{fig:comp_cap_n1020TG}. For $1.68 \times 10^4 < N \le 1.17 \times 10^5$, our Alg.~\ref{alg:nwaysort} has the same number of gates as the SS-Mk. For $N > 1.17 \times 10^5$, the SS-Mk in \cite{gao1997sloping} needs fewer gates.
For $n_b = 20$, we have similar results.
For $N \le 3.71 \times 10^5$, our Alg.~\ref{alg:nwaysort} has fewer or the same number of gates than the SS-Mk. For some segments in $3.71 \times 10^5 < N \le 2.47 \times 10^6$, our Alg.~\ref{alg:nwaysort} has fewer gates than the SS-Mk. For $N > 2.47 \times 10^6$, the SS-Mk in \cite{gao1997sloping} needs fewer gates.

\begin{figure}[!t]
\centering
\includegraphics[width=8.5cm]{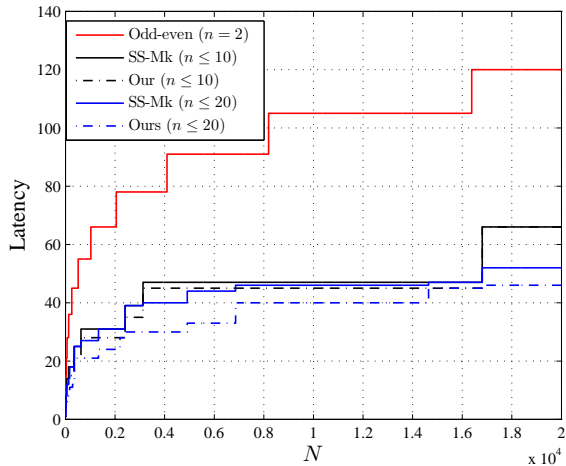}
\caption{Comparison of the latency ($n\le 10$ and $n\le 20$) for sorting $N$ inputs via the SS-Mk in \cite{gao1997sloping} and our Alg.~\ref{alg:nwaysort}.}
\label{fig:latency_cap_n1020TG}
\end{figure}

Similarly, we compare the latency of our sorting algorithm with the SS-Mk in \cite{gao1997sloping}. The latencies are obtained by plugging the corresponding optimal $p$'s into Eqs.~(\ref{eq:Gour_L}) and (\ref{eq:GssMk_L}) and shown in Fig.~\ref{fig:latency_cap_n1020TG} for $N \le 2 \times 10^4$.
Note that the minimization of the number of gates is essentially to minimize the latency, since each $N$ is fixed in Eqs.~(\ref{eq:Gour_L}) and (\ref{eq:GssMk_L}).
Fig.~\ref{fig:latency_cap_n1020TG} also shows the minimal latencies of the Batcher's odd-even algorithm. All the latencies are illustrated by staircase curves. From Fig.~\ref{fig:comp_cap_n1020TG}, the Batcher's odd-even algorithm using 2-sorters has the largest latency.
For both our Alg.~\ref{alg:nwaysort} and the SS-Mk \cite{gao1997sloping}, the latency can be reduced by having a larger $n_b$.
The latency of our Alg.~\ref{alg:nwaysort} is not greater than the SS-Mk for $N \le 2 \times 10^4$ for both $n_b = 10$ and $n_b = 20$ as shown in Fig.~\ref{fig:latency_cap_n1020TG}. This is because our Alg.~\ref{alg:nwaysort} tends to use large basic sorters, leading to less stages.

We also compare the number of gates with buffers for $N$ being a power of two.
The numbers of gates are minimized by varying $p$ according to Eqs.~(\ref{eq:Gour_L}) and (\ref{eq:GssMk_L}) for our algorithm and the SS-Mk \cite{gao1997sloping}. Note the optimal $p$'s are different from those in Sec.~\ref{sec:comp}.
The results are shown in Table~\ref{tab:comp_n2010TG}, where columns two to four show the numbers of gates for the SS-Mk, our Alg.~\ref{alg:nwaysort}, and the reduction of our Alg.~\ref{alg:nwaysort}, respectively, with $n\le 20$, and columns five to seven show those with $n\le 10$. For $n\le 10$ and $n\le 20$, there are up to 25\% and 39\% fewer gates, respectively, than the SS-Mk in \cite{gao1997sloping} for $N = 2^i$ with $i=1,5,\cdots,16$.
It is observed that fewer and the same number of gates are needed for $n\le 20$ than for $n\le 10$ for all $N=2^i$ with $i=1,2,\cdots,16$. The reduction percentage of $n\le 20$ is also greater than or equal to that of $n\le 10$ for all $N=2^i$ with $i=1,2,\cdots,16$ but $N=16$.
This means our sorting network takes better advantage of larger basic sorters.

\begin{table}[!t]
\caption{Comparison of the number of gates with buffers for sorting $N=2^k$ inputs ($1\le k \le 16$) with $n\le 20$ via the SS-Mk in \cite{gao1997sloping} and our Alg.~\ref{alg:nwaysort}.}\label{tab:comp_n2010TG}
\begin{center}
\begin{tabular}{|c|c|c|c|c|c|c|}
\hline
\multirow{3}{*}{$N$} & \multicolumn{3}{c|}{$n\le 20$} & \multicolumn{3}{c|}{$n\le 10$}\\
\cline{2-7}
& \multirow{2}{*}{SS-Mk} & \multirow{2}{*}{Ours} & Rd. & \multirow{2}{*}{SS-Mk} & \multirow{2}{*}{Ours} & Rd. \\
& & & (\%) & & & (\%)\\
\hline
$2$	&	1   &   1	&	0.00	    &	1	 &	 1  &   0.00\\
\hline
$2^2$	&	4   &   4	&	0.00		&	4    &   4	&	0.00\\
\hline
$2^3$	&	8   &     8	&	0.00	    &	32   &   32	&	0.00\\
\hline
$2^4$	&	16  &   16	&	0.00		&	96	&	 80  &    16.67\\
\hline
$2^5$	&	256  &   192	&	25.00		&	256 & 192	&	25.00\\
\hline
$2^6$	&	768    &   512	&	33.33		&	896	&	768 & 14.29\\
\hline
$2^7$	&	1792   &   1152	&	35.71		&	2304	&	1920 & 16.67\\
\hline
$2^8$	&	4608	&    2816	&	38.89	&	4608	&	3840 & 16.67	\\
\hline
$2^9$	&	12800   &   11264	&	12.00		&	12800	&	11264    & 12.00\\
\hline
$2^{10}$	&	27648   &   21504	&	22.22		&	31744	&	28672 & 9.68\\
\hline
$2^{11}$	&	63488	&   49152	&	22.58 	&	63488	&	57344 & 9.68\\
\hline
$2^{12}$	&	163840	&   122880	&	25.00 	&	192512	&	184320 & 4.26\\
\hline
$2^{13}$	&	376832	&   327680	&	13.04	&	385024	&	368640 & 4.26\\
\hline
$2^{14}$   &   770048  &   737280  &   4.26   &   770048  &  737280 &  4.26\\
\hline
$2^{15}$   &   2162688  &   1900544  &   12.12   &   2162688  &  2162688 &  0.00 \\
\hline
$2^{16}$   &   4325376  &   3801088  &   12.12   &   4325376  &  4325376 &  0.00 \\
\hline
\end{tabular}
\end{center}
\end{table}

For $N$ being a power of prime, we compare the number of gates without buffers according to Eqs.~(\ref{eq:GBour_L}) and (\ref{eq:GBssMk_L}).
For $N\le 3\times 10^4$, we search for the same $N$'s for our Alg.~\ref{alg:nwaysort} and the SS-Mk with the minimum number of gates. The results are shown in Tables~\ref{tab:comp_n10TGwoBuf} and \ref{tab:comp_n20TGwoBuf} for $n\le 10$ and $n\le 20$, respectively, where columns three and four show the numbers of gates for the SS-Mk and our Alg.~\ref{alg:nwaysort}, and column five shows the reduction of our Alg.~\ref{alg:nwaysort}.
For all $N$'s except for $N=7^5$, our Alg.~\ref{alg:nwaysort} has no more gates than the SS-Mk in \cite{gao1997sloping}.
There are up to 13\% and 23\% fewer gates than the SS-Mk in \cite{gao1997sloping} for $n\le 10$ and $n\le 20$, respectively. This means our sorting network takes better advantage of larger basic sorters.
We also remark that using a larger sorter size $n$ may reduce the number of gates for sorting $N=n^p$ inputs.
For all common $N$'s for $n\le 10$ in Table~\ref{tab:comp_n10TGwoBuf} and $n\le 20$ in Table~\ref{tab:comp_n20TGwoBuf}, the same number of gates is needed, since the same sorter size $n$ is used.
For all remaining $N$'s except for $N=3^9$ in Table~\ref{tab:comp_n10TGwoBuf}, there is a  corresponding larger $N$'s in Table~\ref{tab:comp_n20TGwoBuf} with fewer gates.
For $N=3^9=19683$ in Table~\ref{tab:comp_n10TGwoBuf} and $N=13^4=28561$ in Table~\ref{tab:comp_n20TGwoBuf}, the latter has about 1\% more gates than the former, but accounts for 45\% more inputs.

\begin{table}[!t]
\caption{Comparison of the number of gates without buffers for sorting $N=n^p$ inputs for $n\le 10$ via the SS-Mk in \cite{gao1997sloping} and our Alg.~\ref{alg:nwaysort}.}\label{tab:comp_n10TGwoBuf}
\begin{center}
\scalebox{0.95}{
\begin{tabular}{|c|c|c|c|c|}
\hline
\multirow{2}{*}{$N$} & \multirow{2}{*}{$n^p$} & \multicolumn{3}{c|}{$n\le 10$} \\
\cline{3-5}
& & \multirow{1}{*}{SS-Mk} & \multirow{1}{*}{Ours} & Rd. (\%)\\
\hline
$2$ & $2$	    &	2   &   2	&	0.00	    \\
\hline
$3$ & $3$	    &	3   &   3	&	0.00		\\
\hline
$5$ & $5$	    &	5   &   5	&	0.00	    \\
\hline
$7$ & $7$	    &	7   &   7	&	0.00		\\
\hline
$9$ & $3^2$   &   29	&	29    & 0.00\\
\hline
$25$ & $5^2$   &   118	&	110 & 6.78\\
\hline
$27$ & $3^3$  &   197	&	188 & 4.57\\
\hline
$49$ & $7^2$   &   305	&	269 & 11.80\\
\hline
$81$ & $3^4$   &   1067 &	998 & 6.47\\
\hline
$125$ & $5^3$    &   1450  &  1315 &  9.31\\
\hline
$128$ & $2^7$    &   2942  &  2942 &  0.00 \\
\hline
$343$ & $7^3$    &   5072  &  4728 &  6.78 \\
\hline
$625$ & $5^4$   &   13489  &  12140 &  10.00 \\
\hline
$729$ & $3^6$   &   22801  &  20411 &  10.48 \\
\hline
$1024$ & $2^{10}$ &  48126  &  48126 &  0.00 \\
\hline
$2401$ & $7^4$   &   63354  &  62254 &  1.74 \\
\hline
$3125$ & $5^5$   &   108175  &  97265 &  10.09 \\
\hline
$4096$ & $2^{12}$ & 278526  &  278526 &  0.00 \\
\hline
$6561$ & $3^8$ &   377375  &  330236 &  12.49 \\
\hline
$8192$ & $2^{13}$ & 655358  &  655358 &  0.00 \\
\hline
$16807$ & $7^{5}$ & 688713  &  704693 &  -2.32 \\
\hline
$19683$ & $3^9$ &   1443791  &  1259711 &  12.75 \\
\hline
\end{tabular}
}
\end{center}
\end{table}

\begin{table}[!t]
\caption{Comparison of the number of gates without buffers for sorting $N=n^p$ inputs for $n\le 20$ via the SS-Mk in \cite{gao1997sloping} and our Alg.~\ref{alg:nwaysort}.}\label{tab:comp_n20TGwoBuf}
\begin{center}
\scalebox{0.95}{
\begin{tabular}{|c|c|c|c|c|}
\hline
\multirow{2}{*}{$N$} & \multirow{2}{*}{$n^p$} & \multicolumn{3}{c|}{$n\le 20$} \\
\cline{3-5}
& & \multirow{1}{*}{SS-Mk} & \multirow{1}{*}{Ours} & Rd. (\%)\\
\hline
$2$ & $2$	    &	2   &   2	&	0.00	    \\
\hline
$3$ & $3$	    &	3   &   3	&	0.00		\\
\hline
$5$ & $5$	    &	5   &   5	&	0.00	    \\
\hline
$7$ & $7$	    &	7   &   7	&	0.00		\\
\hline
$11$ & $11$	&	11  &   11	&	0.00		\\
\hline
$13$ & $13$	&	13    &   13	&	0.00		\\
\hline
$17$ & $17$	&	17   &   17	     &	0.00         \\
\hline
$19$ & $19$	&	19	&    19	      &	0.00         \\
\hline
$25$ & $5^2$	&	118   &   110	&	6.78		 \\
\hline
$27$ & $3^{3}$	&	197   &   188	&	4.57		 \\
\hline
$49$ & $7^{2}$	&	305	&   269	      &	11.80 	     \\
\hline
$121$ & $11^{2}$	&	1117	&   917	&	17.91 	 \\
\hline
$125$ & $5^{3}$	&	1450	&   1315	&	9.31	 \\
\hline
$169$ & $13^{2}$   &   1814  &   1454  &   19.85     \\
\hline
$289$ & $17^{2}$   &   3970  &   3074  &   22.57     \\
\hline
$361$ & $19^{2}$   &   5501  &   4205  &   23.56     \\
\hline
$625$ & $5^{4}$   &   13489  &   12140  &   10.00    \\
\hline
$729$ & $3^{6}$   &   22801  &   20411  &   10.48    \\
\hline
$1331$ & $11^{3}$   &   29107  &   26668  &   8.38    \\
\hline
$2197$ & $13^{3}$   &   54703  &   50763  &   7.20    \\
\hline
$2401$ & $7^{4}$   &   63354  &   62254  &   1.74     \\
\hline
$3125$ & $5^{5}$   &   108175  &   97265  &   10.09   \\
\hline
$4913$ & $17^{3}$   &   156812  &   143443  &   8.53  \\
\hline
$6859$ & $19^{3}$   &   239590  &   221052  &   7.74  \\
\hline
$14641$ & $11^{4}$   &   564513  &   562214  &   0.41  \\
\hline
$16807$ & $7^{5}$   &   688713  &   704693 &  -2.32 \\
\hline
$28561$ & $13^{4}$  &   1230724 &   1271788 &   -3.34\\
\hline
\end{tabular}
}
\end{center}
\end{table}

\section{Conclusion}
\label{sec:conclusion}
In this work, we proposed a new merging algorithm based on $n$-sorters for parallel sorting networks, where $n$ is prime. Based on the $n$-way merging, we also proposed a merge sorting algorithm. Our sorting algorithm is a direct generalization of odd-even merge sort with $n$-sorters as basic blocks.
By using larger sorters ($2 \le n \le 20$), the number of sorters as well as the latency is reduced greatly. In comparison with other multiway sorting networks in \cite{gao1997sloping}, our implementation has a smaller latency and fewer sorters for wide ranges of $N\le 1.46\times 10^4$. We also showed an application of sorting networks implemented by linearly scaling sorters in threshold logic and have a similar conclusion that the number of gates can be greatly reduced by using larger sorters.

\appendices
\section{Proofs}
\subsection{Proof of Lemma~\ref{lm:subgroupsort}}
\label{pf:subgroupsort}
\begin{proof}
  The proof of the lemma can be reduced to showing that for $l > 0$ any two wires $s, s+l \in \mathbb{Z}_m$ of each list are sorted as shown in Fig.~\ref{fig:mnsort}(b). We prove the lemma by contradiction.
  The inputs satisfy $x_{j,s} \le x_{j,s+l}$ for $j \in \mathbb{Z}_n$ and $s, s+l \in \mathbb{Z}_m$.
  Suppose there exist $k \in \mathbb{Z}_n$ and $s, s+l \in \mathbb{Z}_m$ such that $x'_{k,s} > x'_{k,s+l}$. Since the sorter for $x_{k,s}$ $k \in \mathbb{Z}_m$ acts as a permutation of the index $k$, we denote such permutation of the sorter connecting wire $s$ as $f: \{1,\cdots,n\} \rightarrow \{1,\cdots,n\}$. Because $f$ is bijection, an inverse $f^{-1}$ exists. Then we have $x_{f^{-1}(t),s+l} \ge^a x_{f^{-1}(t),s} = x'_{t,s} \ge x'_{k,s} > x'_{k,s+l}$ for $k \le t \le n$, where the ``$\ge^a$'' is because the inputs are sorted and the ``$=$'' is due to the permutation.
  There are $n-k+1$ inputs of $x_{f^{-1}(t),s+l}$ satisfying $x_{f^{-1}(t),s+l} > x'_{k,s+l}$. However, at most $n-k$ outputs satisfy $x'_{t,s+l} > x'_{k,s+l}$ for $t \in \{k+1,k+2,\cdots,n\}$, resulting in a contradiction. Hence, all lists are self sorted after applying $n$-sorters.
\end{proof}

\subsection{Proof of Lemma~\ref{lm:sorter4case}}
\label{pf:sorter4case}
\begin{proof}
 First, we show that the first connections of adjacent two sorters belong to either the same list or adjacent two lists. Let $(j,t_1)$ and $(j+l,t_2)$ be the first connections of adjacent two sorters $S1$ and $S2$, respectively, where $(j,t)$ denotes wire $t$ in list $j$.
 If $l>1$, the connection of $S1$ in list $j+l-1$ should be wire $m$; otherwise, $S2$ would have a valid connection in list $j+l$. For lists $j$ to $j+l-2$, only wires $m$ in each list are connected by $S1$, since wire $m$ can be connected to the preceding list only by a $(m-1)$-spaced sorter. Hence, $S1$ is the last $(m-1)$-spaced sorter in stage 1 and $S2$ does not exist.
 Similarly, we can show that the last connections of adjacent two sorters $S1$ and $S2$ belong to either the same list or adjacent two lists. This gives us a total of four cases as shown in Fig.~\ref{fig:m2sort}, where $b\ge a+1$ for Fig.~\ref{fig:m2sort}(a)-(c), and $b\ge a$ for Fig.~\ref{fig:m2sort}(d) such that $S1$ and $S2$ have a size of at least two.

 If $m$ is prime, no adjacent two sorters belong to case IV, which is equivalent to showing that $m$ is a composite number if case IV in Fig.~\ref{fig:m2sort} exists. Assume two adjacent sorters $S1$ and $S2$ belong to case IV. Let the first connection of $S1$ be $(j,m)$ and the last connection of $S2$ be $(j+p, 1)$. The last connection of $S1$ satisfies $(k+1)p \equiv 0 \quad \mbox{mod }m$. We have $m \mid (k+1)p$. Since case IV is not possible in the first stage, we have $p <m$. Since two adjacent sorters connect two adjacent wires in at least one list, we have $p > 1$.
 If $k=0$, $S1$ would connect the last and first wires of adjacent lists, respectively, in which case $S2$ does not exist. We have $1 < k+1 < m$.
 So $m$ should have a proper factor dividing $k+1$ or $p$. Hence, $m$ is a composite number.
\end{proof}

\subsection{Proof of Theorem~\ref{thm:groupsort}}
\label{pf:groupsort}
\begin{proof}
  The theorem can be proved by induction on $i$.
  In stage 1, $m$-sorters are applied on corresponding wires of all $m$ lists. According to Lemma~\ref{lm:subgroupsort}, the outputs of each list are sorted.
  Assume any two adjacent wires $s$ and $s+1$ in list $j$ are sorted after stage $i-1$, $x^{(i-1)}_{j,s} \le x^{(i-1)}_{j,s+1}$ for $1\le j \le n$ and $1\le s \le m-1$. We will show that $x^{(i)}_{j,s} \le x^{(i)}_{j,s+1}$ for $1\le j \le n$ and $1\le s \le m-1$.

  According to Lemma~\ref{lm:sorter4case}, for a prime $m$, there are three cases of two adjacent sorters $S1$ and $S2$ as shown in Fig.~\ref{fig:m2sort}(a)-(c).
  \begin{enumerate}
    \item For case I, let $y^{(i-1)}_{j,1}$ and $y^{(i-1)}_{j,2}$ be the two adjacent wires in list $j$ connected by adjacent two sorters in stage $i-1$ for $a\le j \le b$. According to Lemma~\ref{lm:subgroupsort} ($n=2$), the outputs of each list are sorted.
    \item For case II, there is an additional single wire $y^{(i-1)}_{b+1}$ connected by $S2$. If $y^{(i-1)}_{b+1,1}=1$, we have $y^{(i)}_{b+1,1}=1$. The last connection of $S2$ can be removed without changing the order of others in $S2$.
        $S1$ and the revised $S2$ reduce to case I and the outputs are sorted according to Lemma~\ref{lm:subgroupsort}. If $y^{(i-1)}_{b+1,1}=0$, we have $y^{(i-1)}_{b,1}=0$. This is because they are connected by the same sorter in stage $i-1$. Then, we have $y^{(i)}_{a,1}=y^{(i)}_{a,2}=0$, which are sorted outputs in list $a$. Remove $y^{(i-1)}_{b+1,1}, y^{(i-1)}_{b,1}, y^{(i)}_{a,1}$, and $y^{(i)}_{a,2}$, the remaining of $S1$ and $S2$ reduce to a smaller configuration of case II. With recursively applying the above approach, $S1$ and $S2$ either reduce to a smaller case I or a single wire, both of which gives sorted outputs.
    \item For case III, there is an additional single wire $y^{(i-1)}_{a-1,m}$ connected by the first sorter. Similarly, the two sorters can be reduced to either a case I or a smaller configuration of case III and the outputs of two adjacent wires in each list are sorted.
  \end{enumerate}
  Assume all lists are self-sorted after stage $i-1$, we have $x^{(i-1)}_{j,1} \le \cdots \le x^{(i-1)}_{j,m}$ for $1 \le j \le n$.
  For stage $1 \le i \le \lceil \frac{m}{2} \rceil$, all wires in lists $j=2,\cdots,n-1$ have connections with some sorters.
  We have $ x^{(i)}_{j,k} \le x^{(i)}_{j,k}$ for $j=2,\cdots,n-1$ and $k=1,\cdots,m-1$. Hence, lists $j=2,\cdots,n-1$ are self-sorted after stage $i$.
  For list 1, $x^{(i-1)}_{1,i-1} \le x^{(i-1)}_{2,1}$ and $x^{(i-1)}_{1,i-1} \le x^{(i-1)}_{1,i}$, we have $x^{(i)}_{1,i-1} \le x^{(i)}_{1,i}$. We have $\langle x^{(i)}_{1,1}, x^{(i)}_{1,2},\cdots, x^{(i)}_{1,i-1} \rangle$, since list 1 is self-sorted after stage $i-1$ and $x^{(i-1)}_{1,k} = x^{(i)}_{1,k}$ for $k=1,\cdots,i-1$. We also have $x^{(i)}_{1,i}, x^{(i)}_{1,i+1}, \cdots, x^{(i)}_{1,m} \rangle$. Hence, list 1 is self-sorted after stage $i$, $x^{(i)}_{1,1}, x^{(i)}_{1,i+1}, \cdots, x^{(i)}_{1,m} \rangle$. Due to symmetry, list $n$ is also self-sorted after stage $i$, $x^{(i)}_{m,1}, x^{(i)}_{m,i+1}, \cdots, x^{(i)}_{n,m} \rangle$.

  To prove that the outputs of $n$ sorted lists $\langle x^{(\lceil \frac{m}{2} \rceil)}_{j,1}, \cdots, x^{(\lceil \frac{m}{2} \rceil)}_{j,m} \rangle$ for $j=1,\cdots,n$ after stage $\lceil \frac{m}{2} \rceil$ are combined as a single sorted list in stage $\lceil \frac{m}{2} + 1 \rceil$, we need to show that $x^{(\lceil \frac{m}{2} \rceil + 1)}_{j,\frac{m+1}{2}} \le x^{(\lceil \frac{m}{2} \rceil + 1)}_{j,\frac{m+1}{2}+1}$ for $j=1,\cdots,n-1$ and $x^{(\lceil \frac{m}{2} \rceil + 1)}_{j,\frac{m+1}{2}-1} \le x^{(\lceil \frac{m}{2} \rceil + 1)}_{j,\frac{m+1}{2}}$ for $j=2,\cdots,n$. Since $x^{(\lceil \frac{m}{2} \rceil)}_{j,\frac{m+1}{2}} \le x^{(\lceil \frac{m}{2} \rceil)}_{j,\frac{m+1}{2}+1}$ and $x^{(\lceil \frac{m}{2} \rceil)}_{j,\frac{m+1}{2}} \le x^{(\lceil \frac{m}{2} \rceil)}_{j+1,1}$, we have $x^{(\lceil \frac{m}{2} \rceil + 1)}_{j,\frac{m+1}{2}} \le x^{(\lceil \frac{m}{2} \rceil+1)}_{j,\frac{m+1}{2}+1}$ for $j=1,\cdots,n-1$. Similarly, we have $x^{(\lceil \frac{m}{2} \rceil + 1)}_{j,\frac{m+1}{2}-1} \le x^{(\lceil \frac{m}{2} \rceil+1)}_{j,\frac{m+1}{2}}$ for $j=2,\cdots,n$
\end{proof}

\subsection{Proof of Lemma~\ref{lm:mergen0s}}
\label{pf:mergen0s}
\begin{proof}
  In stage $i-1$, there are $n^{i-1}$ sorted lists of $n$ values with respect to each $q$ ($q=1,\cdots,n^{p-i}$).
  Since the outputs of each merging network are sorted after stage $i-1$, we can replace each merging network by an $n^i$-sorter.
  According to Lemma~\ref{lm:subgroupsort}, the outputs of each new formed list after stage $i$ are sorted, $x^{(i)}_{j,q} \le x^{(i)}_{j,n^{p-i-1} + q} \le x^{(i)}_{j,(n-1)n^{p-i-1} + q}$ for $j=1,\cdots,n^i$.
  Since the corresponding wires in the new lists are connected by the same $n^i$-sorter in stage $i-1$, we have $x^{(i)}_{j,q} \le x^{(i)}_{j+1,q}$ for $j=1,\cdots, n^i-1$. Hence, $r^{(i)}_{j,q} \ge r^{(i)}_{j+1,q}$ for $j=1,\cdots, n^i-1$.

  For $r^{(i)}_{s,q} = n > r^{(i)}_{s+1,q} \ge \cdots \ge r^{(i)}_{s+l,q} > 0 = r^{(i)}_{s+l+1,q} \quad \mbox{for} \quad l \le n$, it is equivalent to prove that $x^{(i)}_{j+n,q} = 1$ if $x^{(i)}_{j,(n-1)n^{p-i-1}+q} = 1$ for $j\in \{1,\cdots,n^i-n\}$.
  For any $q \in \{1,\cdots, n^{p-1-i}\}$ in stage $i$, there are $n^i$ lists of $n$ values. Suppose $x^{(i)}_{j,(n-1)n^{p-i-1}+q} = 0$ for $j\le s$ and $x^{(i)}_{s+1,(n-1)n^{p-i-1}+q} = 1$.
  If $t$ ($t\le s$) zeros of $x^{(i)}_{j,(n-1)n^{p-i-1}+q}$ are from the same list of the original $n$ sorted lists, there are at most $t+1$ zeros of $x^{(i)}_{j,q}$ from that same list. Since $x^{(i)}_{j,(n-1)n^{p-i-1}+q}=0$ for $j\le s$ are from at most $n$ original lists, there are at most $s+n$ zeros in $x^{(i)}_{j,q}$, implying that $x^{(i)}_{s+n,q}=1$.
  Hence, $x^{(i)}_{j+n,q} = 1$ if $x^{(i)}_{j,(n-1)n^{p-i-1}+q} = 1$ for $j\in \{1,\cdots,n^i-n\}$.
\end{proof}

\subsection{Proof of Theorem~\ref{thm:nwaymerge}}
\label{pf:nwaymerge}
\begin{proof}
  In stage 1, all outputs with respect to the operation of the same Alg.~\ref{alg:nway} are sorted.
  For any $q\in \{1,\cdots, n^{p-1-i} \}$ in stage $i$, according to Lemma~\ref{lm:mergen0s}, at most $n$ consecutive lists are not full of zeros. All preceding lists are all-zero lists and all following lists are all-one lists.
  Hence, the combining network in stage $i$ is to sort $n$ lists of $n$ values, which is reduced to Alg.~\ref{alg:nway}. In stage $p-1$, we have $q=1$ and the single sorted list, $\langle x^{(i)}_{1,q}, x^{(i)}_{1,n^{p-i-1} + q}, \cdots, x^{(i)}_{1,(n-1)n^{p-i-1} + q}, x^{(i)}_{2,q}, x^{(i)}_{2,n^{p-i-1} + q},$ $\cdots, x^{(i)}_{2,(n-1)n^{p-i-1} + q}, \cdots, x^{(i)}_{n^i,q}, x^{(i)}_{n^i,n^{p-i-1} + q}, \cdots$, $x^{(i)}_{n^i,(n-1)n^{p-i-1} + q} \rangle$, contains $n^p$ values, implying all inputs are sorted as a single list.
\end{proof}

\bibliographystyle{IEEEtran}
\bibliography{TL}

\end{document}